# Many-body calculations for periodic materials

# via quantum machine learning

Shu Kanno[1], Tomofumi Tada[1,2]*

[1]*Materials Research Center for Element Strategy, Tokyo Institute of Technology, 4259 Nagatsuta, Midori-ku, 226-8501 Yokohama, Japan*

[2]*Kyushu University Platform of Inter/Transdisciplinary Energy Research, Kyushu University, 744 Motooka, Nishi-ku, 819-0395 Fukuoka, Japan*

* tada.tomofumi.054@m.kyushu-u.ac.jp

# Abstract

A state-of-the-art method that combines a quantum computational algorithm and machine learning, so-called quantum machine learning, can be a powerful approach for solving quantum many-body problems. However, the research scope in the field was mainly limited to organic molecules and simple lattice models. Here, we propose a workflow of quantum machine learning applications for periodic systems on the basis of an effective model construction from first principles. The band structures of the Hubbard model of graphene with the mean-field approximation are calculated as a benchmark, and the calculated eigenvalues show good agreement with the exact diagonalization results within a few meV by employing the transfer learning technique in quantum machine learning. The results show that the present computational scheme has the potential to solve many-body problems quickly and correctly for periodic systems using a quantum computer.

# I. INTRODUCTION

Quantum many-body effects lead to various exotic phenomena that cannot be interpreted in the independent electron picture owing to electron correlations. Typical physical properties induced by the quantum many-body effect include magnetism and superconductivity, which have been confirmed in inorganic material, e.g., the strong magnetism of neodymium compounds and the high-temperature superconductivity of copper oxides. Quantum many-body problems are related to quantum many-body effects. Thus, correctly solving many-body problems in the case of inorganic materials is important for accelerating correlated materials exploration. Many inorganic compounds have a structural periodicity, and these atomic/ionic configurations are identified as crystal structures. In a periodic system, the interactions characterizing the system are distributed over a wide range through its periodicity, and energy levels are not isolated as in molecules but are continuous bands specified by the wave vector $\boldsymbol{k}$. Thus, the computational cost for periodic systems far exceeds that for isolated ones. Consequently, the electronic structure calculation for inorganic crystals has mainly been performed using density functional theory (DFT). DFT can more easily incorporate electron correlations than wave function methods such as Hartree-Fock and configuration interaction because the correlation effects can be included through the effective potential as approximated forms. In DFT, a uniform electron gas-based approximation such as the local density approximation (LDA) or generalized gradient approximation (GGA) is often used to represent the electron correlation energy. However, in some transition metal compounds, these approximations might not be justified because of a strong electron correlation leading to electron localization. The electronic states of transition metal oxides including cuprate superconductors cannot be described accurately in DFT, especially near the Fermi energy, which in turn results in wrong predictions for physical properties.

Effective many-body Hamiltonian construction through the downfolding method [1] has been proposed to increase the predictability of many-body problems, which corresponds to a beyond-DFT method. In this method, after band structure calculation using a standard method such as DFT, the many-body contributions from high-energy regions/bands with respect to the Fermi level are renormalized into the low-energy bands near the Fermi level (i.e., downfolding). This can rewrite the eigenvalue problem of quantum many-body systems represented by many orbitals into smaller-sized effective models written by few orbitals. A well-known example is the Hubbard model. When this model is constructed accurately and solved correctly, it is possible to predict the correct physical properties even for inorganic materials showing strong electron correlation. For example, the magnetic properties of iron-based superconductors [2] and the superconducting properties of a cuprate superconductor [3] have been calculated fairly accurately.

While eigenvalue problems based on the many-body effective models have been solved by variational Monte Carlo methods [2] and dynamical mean-field theory [4], highly accurate approaches using machine learning have been proposed in recent years [5]. One widely adopted approach involves a restricted Boltzmann machine (RBM) [6], which is a two-layered neural network constructed by interconnected visible and hidden layers. RBM can be broadly characterized in terms of two parameters: the node bias and the coupling strength between nodes. Each layer consists of "units" that can take a binary state, 1 or -1, which corresponds to the electron spin state of up or down in a physical system. The units in the visible layer correspond to the physical sites defined in a many-body effective model, and the electron interaction is expressed by the units in the hidden layer. Machine learning using a neural network such as RBM is potentially an effective approach to solving the quantum many-body problem accurately, with one serious disadvantage, namely, that the computational cost would increase exponentially with the number of electrons.

In recent years, "quantum algorithms," which greatly accelerate the computational time for specific problems, have attracted much attention as a novel computational technique. Since quantum algorithms are employed through a superposition state in a quantum computer, it is expected that their representation ability, particularly for quantum states, will be much higher, and the computational time extremely reduced as compared to conventional computational algorithms, so-called classical algorithms. As a future application of the quantum algorithm for many-body problems, attempts at combining quantum algorithms and machine learning (i.e., quantum machine learning) have been made. For example, topological phase detection [7] and electronic structure calculations for ground states [8] have been studied by quantum machine learning. However, the models to be solved with quantum machine learning have so far been limited to small molecules or simple lattice models, with no applications for inorganic periodic materials. An expansion of achievable targets in quantum machine learning would accelerate basic research in quantum computation. In this paper, we present a three-step workflow to solve the many-body problem in periodic systems such as inorganic materials by using quantum machine learning from first-principles calculation. The workflow is applied to the mean-field Hubbard model of graphene.

The overall computational workflow proposed in this study is shown in Fig. 1: (Step 1) DFT is adopted to calculate global electronic structures including high-energy bands, and target bands for the many-body problem are identified; (Step 2) an effective model for the many-body problem is constructed by using an effective on-site Coulomb repulsion for the target bands, and (Step 3) quantum machine learning is executed for the eigenvalue problem defined with this effective model. Physical properties can be calculated after quantum machine learning. When the system size is small, an approach considering all electrons may be practical, but for larger systems it will be impractical because the number of qubits and gate operations increases with system size. This is also the case for a system containing heavy elements in which the principal quantum number of the elements is large. In the present study, the number of qubits is suppressed to a small number by using an effective model constructed with only the target orbitals [9]. Thus, quantum machine learning can be executed for inorganic crystals. Since all computational methods in this workflow (i.e., DFT and effective model construction) can be replaced with other methods, various target models can be calculated in the present scheme depending on the purpose. For example, effective models can be obtained through the constrained random phase approximation (cRPA) [10], the linear response method [11], and the constrained local density approximation (cLDA) [12].

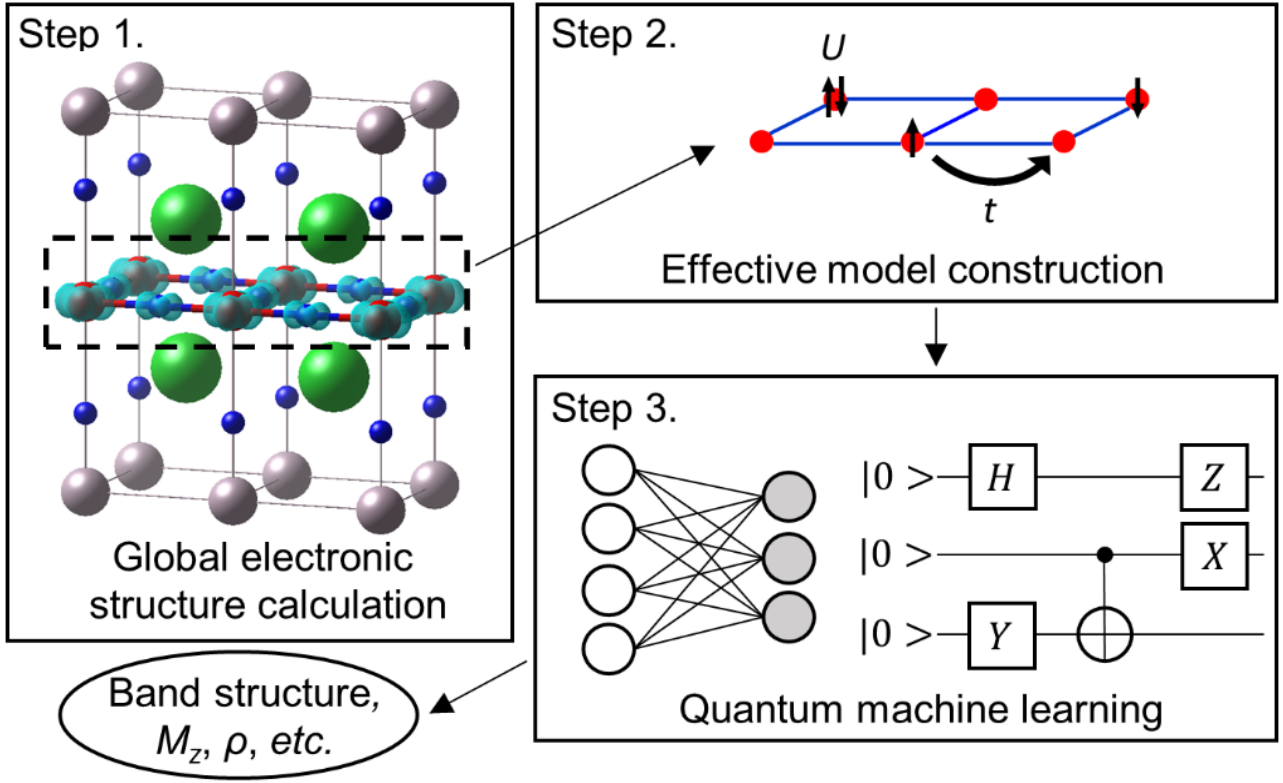

**Fig. 1. Workflow for calculation of materials properties using quantum algorithm for many-body problems of inorganic systems.** The workflow comprises three steps: (Step 1) global electronic structure calculation including high-energy region, (Step 2) effective model construction of target orbitals, (Step 3) quantum machine learning using the model Hamiltonian. At the end of the workflow, physical properties such as band structure, magnetism, electrical conductivity, *etc.*, can be obtained.

Let us outline each method in the workflow. The model details and calculation conditions are described in Supplementary Materials.

Step 1. The global electronic structure is calculated using the crystal structure of the target. We selected graphene as the target system in the present study, and adopted DFT for the global band structure calculations. Graphene is a two-dimensional (2D) periodic system in which the unit cell is composed of two carbon atoms. Target bands are extracted from the calculated band structure.

Step 2. An effective model is constructed for the target band. In graphene, the target bands are composed of $2p_z$ orbitals of two carbon atoms, i.e., π bands. The maximally localized Wannier function [13] is adopted for basis construction for the effective model, and a Hubbard model for the target bands is constructed:

$$H = \sum_{iji'j'}^{A,B} \sum_{ll'} \sum_{\sigma}^{\uparrow,\downarrow} t_{i'j'l';ijl} a^\dagger_{i'j'l'\sigma} a_{ijl\sigma} + U \sum_{ij}^{A,B} \sum_{l} n_{ijl\uparrow} n_{ijl\downarrow} = H_{hop} + H_C, \qquad (1)$$

where $i, j, i', j'$ are the lattice site indices, $l, l'$ are atomic sites $A$ and $B$ in the lattice, $\sigma(=\uparrow,\downarrow)$ is the spin index, $a^\dagger/a$ is the creation/annihilation operator, and $n_{ijl\sigma}(= a^\dagger_{ijl\sigma}a_{ijl\sigma})$ is the number operator. $t_{i'j'l';ijl}$ is the hopping energy between site $l$ of lattice $(i,j)$ and site $l'$ of lattice $(i',j')$, and $U$ is the effective on-site Coulomb repulsion energy (Hubbard $U$). The hopping energy $t$ was obtained in the Wannier orbital representation. The value of Hubbard $U$ (= 9.3 eV) was adopted from a previous study [14], in which $U$ was calculated by using the first-principles method with the constrained random-phase approximation (cRPA) [10]. For the sake of simplicity, we replace $H_C$ with $H_C^{mf}$ by using the mean-field approximation.

$$H_C^{mf} = U \sum_{ij} \sum_{l}^{A,B} (n_{ijl\uparrow} <n_\downarrow> + <n_\uparrow> n_{ijl\downarrow} - <n_\uparrow><n_\downarrow>), \quad (2)$$

where $<n_\sigma>$ is the expectation value of the particle number operator of spin $\sigma$. In this study, we applied Fourier transform to the creation/annihilation operators to treat periodic structures as follows:

$$a^\dagger_{ijl\sigma} = \frac{1}{\sqrt{N}} \sum_k e^{i\mathbf{k}\cdot\mathbf{r}_{ijl}} a^\dagger_{kl\sigma} \quad (3)$$

$$a_{ijl\sigma} = \frac{1}{\sqrt{N}} \sum_k e^{-i\mathbf{k}\cdot\mathbf{r}_{ijl}} a_{kl\sigma} \quad (4)$$

$$\mathbf{r}_{ijl} = i\mathbf{a}_1 + j\mathbf{a}_2 + \mathbf{r}_l,$$

where $\mathbf{k}$ is a wave vector, $N$ is the number of unit cells of the crystal, $\mathbf{a}_1$ and $\mathbf{a}_2$ are the lattice vectors, and $\mathbf{r}_l$ (= $\mathbf{r}_A$ or $\mathbf{r}_B$) is a position vector indicating the atomic sites. As a result of the Fourier transform on $H_{hop}$ and $H_C^{mf}$, $H$ can be represented by the sum of the independent $H_k$,

$$H = \sum_k H_k, \quad (5)$$

where $H_k$ is given by

$$H_k = \sum_{l,l'}^{A,B} \sum_\sigma^{\uparrow,\downarrow} \sum_m t_{kl'lm} a^\dagger_{kl'\sigma} a_{kl\sigma} + U \sum_l (n_{kl\uparrow} <n_\downarrow> + <n_\uparrow> n_{kl\downarrow} - <n_\uparrow><n_\downarrow>). \quad (6)$$

$t_{kl'lm}$ is the $m$-th neighboring hopping parameter between sites $l'$ and $l$ in the reciprocal space (see supplemental material for details). In the present work, the mean-field approximation is adopted in the Hubbard Hamiltonian (Eq. (2)) because our main purpose is to validate the feasibility of quantum machine learning for periodic systems. The Hubbard Hamiltonian without a mean-field approximation could be treated without any loss of accuracy; the machine learning approach has already been applied to the Hubbard Hamiltonian for a lattice model [15] with high accuracy. In addition, the combination of quantum algorithms and non-mean-field Hamiltonians for crystalline materials has been reported using the plane-wave technique, although only the number of required quantum gates was estimated (i.e., no demonstration of quantum algorithm) [16].

As a practical form, the creation and annihilation operators of the Hubbard Hamiltonian are transformed into the Pauli operators (spin operators) using the Jordan-Wigner transformation [17] for quantum machine learning; the down spin $|\downarrow\rangle$ corresponds to an occupied state $|1\rangle$, and the up spin $|\uparrow\rangle$ corresponds to an unoccupied state $|0\rangle$ of electrons in the transformed Hamiltonian. In what follows, we will use the notation for occupied/unoccupied states, but apply the notation for spin up/down states when necessary.

Step 3. By applying quantum machine learning to the Hubbard model defined above, the band structures including electron correlation are calculated. The algorithm for quantum machine learning is essentially based on the framework proposed by Xia et al [8]. Details regarding the algorithm can be found elsewhere [8]. Let us now outline the present algorithm to clarify how we modified the quantum machine learning approach to make it applicable to periodic systems such as inorganic crystals. First, the algorithm uses a restricted Boltzmann machine (RBM) to represent the wave function as a linear combination of occupied states such as $|0011\rangle, |0111\rangle, |0100\rangle$, *etc.*, where the four basis functions in each ket vector represent the states A↑, A↓, B↑ and B↓, respectively, and 1/0 denotes the occupied/unoccupied states in the Hubbard Hamiltonian at each k-point (For example, $|0100\rangle$ is $a_{A\downarrow}^{\dagger}|0000\rangle$). Figure 2A illustrates the Boltzmann machine used in this study. It consists of visible and hidden layers with four units in each layer and a complex layer with two units. In previous research [8], the number of units in the complex layer called the "sign layer" was limited to one, and the wave functions were limited to real numbers. By contrast, in the present study, complex numbers can be expressed by the two sites in the complex layer, which are necessary for the wavefunctions of periodic systems. The wave function $|\Psi\rangle$ in spin up/down notation is shown in Eq. (7). P(x) and s(x) are, respectively, the Boltzmann distribution and complex number distribution defined in Eqs. (8) and (9). The spin coordinate is represented by $x = \{\sigma_1 \dots \sigma_4\}$, where the values of $\sigma_i$ (visible) and $h_j$ (hidden) are 1 for spin up and -1 for spin down. In Fig. 2A, $a_i, b_j, c,$ and $e$ are the bias parameters and $d_i, f_i,$ and $w_{ij}$ are the connection parameters, where indices $i$ and $j$ run from 1 to 4.

$$|\Psi\rangle = \sum_x \sqrt{P(x)}\, s(x)|x\rangle \tag{7}$$

$$P(x) = \sum_{\{h_j\}} e^{\sum_i a_i \sigma_i + \sum_j b_j h_j + \sum_{ij} w_{ij} \sigma_i h_j} \tag{8}$$

$$s(x) = \tanh\left(\left(c + \sum_i d_i \sigma_i\right) + i\left(e + \sum_i f_i \sigma_i\right)\right) \tag{9}$$

Normalization factors are omitted in Eqs. (7)-(9) for simplicity. The summation in Eq. (8) (i.e., $\{h_j\}$) corresponds to the trace for all the possible spin configurations of the hidden layer units. After the wave function is obtained by Eq. (7), the expected value $<H_\mathbf{k}>$ for the lowest eigenvalue is calculated by Eq. (10), and the RBM parameters are updated by the conventional optimization method, Adam [18].

$$<H_\mathbf{k}> = \frac{<\Psi|H_k|\Psi>}{<\Psi|\Psi>} \quad (10)$$

The same procedure is iterated until the difference of $<H_\mathbf{k}>$ in an iteration step is less than the threshold value (see supplemental material). When higher eigenvalues need to be calculated, we can use the algorithms for determining excited states [19-21].

The quantum algorithm is used to obtain the Boltzmann distribution by RBM. Other calculation procedures such as optimization of RBM parameters are performed classically. Figure 2B shows the quantum circuit used in this study. The quantum circuit mainly consists of two types of operations: (i) a one-qubit operation, $R_y$, (See the left side of Fig. 2B) that corresponds to a rotational operation whose angle is determined by the bias parameters $a_i$(visible) and $b_j$(hidden), and (ii) a three-qubit operation, $C_1$-$C_2$-$R_y$, that is a controlled-controlled-rotation whose angle is determined by the connection parameter $w_{ij}$ ("$C_1$-$C_2$-$R_y$" of Fig. 2B). After preparing a superposition state composed of all configurations in which the weight of each component depends on the bias parameters $a_i, b_j$ (i.e., the first 1-qubit operation), entanglement states composed of all configurations in which the weight of each component depends on the connection parameters $w_{ij}$ are generated (i.e., 3-qubit operations and measurements). A Boltzmann distribution for all configurations of the visible and hidden layers can be generated through the quantum circuit. The number of the gate operations required for one sampling of this quantum algorithm is O(*mn*) at best; *m* and *n* are the numbers of units in the visible and hidden layers, respectively. The qubits required in the present network structure are 9 (= *m* + *n* + 1 = 4 + 4 + 1) qubits, where the third term, 1, comes from the ancilla qubit in Fig. 2B.

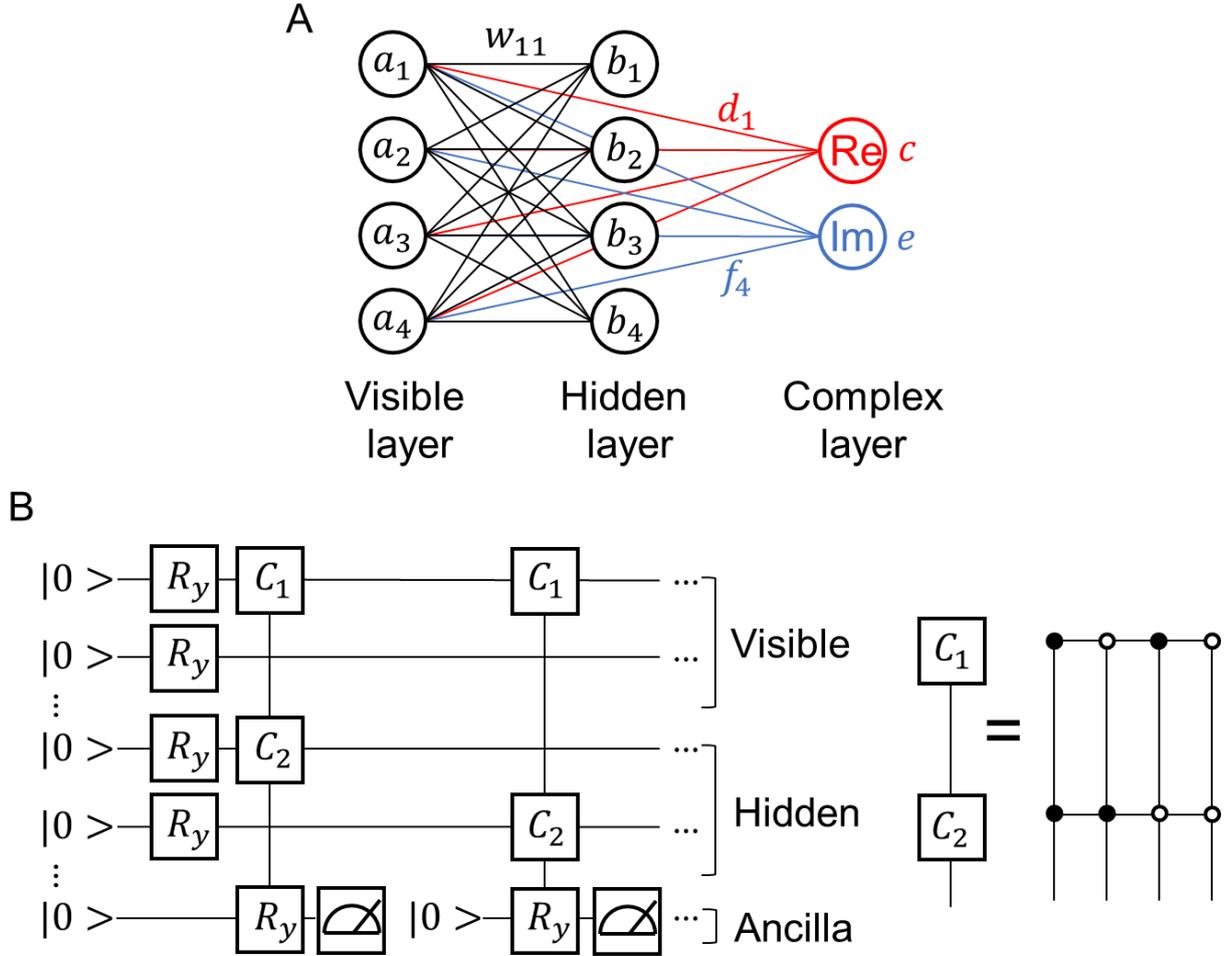

**Fig. 2. RBM and a quantum circuit, generating Boltzmann distribution $P(x)$.** (A) RBM consists of visible and hidden layers with four units in each layer and a complex layer with two units. (B) The quantum circuit consists of 1-qubit operation and 3-qubit operation with ancilla measurement related to the connection parameters. $R_y$ denotes an $R_y$ rotation gate, and $C_1$ and $C_2$ are combinations of four types of control gates (written on the right side of the figure). The ancilla qubit is measured with the basis of $< 1|$ after the $C_1$- $C_2$- $R_y$ operation. (For the rotation angle of each $R_y$ rotation gate, refer to a previous research [8].)

## II. RESULTS

1. Construction of electron correlation model

      Figure 3A shows the energy band structure of graphene calculated with DFT. The energy bands using maximally localized Wannier orbitals corresponding to carbon $p_z$ orbitals are also depicted (the red dashed lines in Fig. 3A). The energy bands with DFT and the Wannier function show a good agreement. Figure 3B shows the crystal structure of graphene and the orbital of the maximally localized Wannier function; a $p_z$-shaped orbital centered at a carbon atom is clearly

confirmed. In this study, for the Hamiltonian treated by quantum machine learning, we adopted hopping terms up to the third nearest neighbor. The maximum difference in energy band level between the truncated and non-truncated cases is about 0.4 eV. In order to construct a many-body Hamiltonian based on the Wannier-based truncated Hamiltonian, we adopted 9.3 eV for $U$ according to a previous theoretical study on graphene [14]. Table 1 lists the on-site/hopping energies, the number of connecting sites, and the distance between the sites. The Hamiltonian $H_\mathbf{k}$ at each k-point in Eq. (6) is created from the obtained $t$ and $U$. The detailed expression is described in the supplemental material.

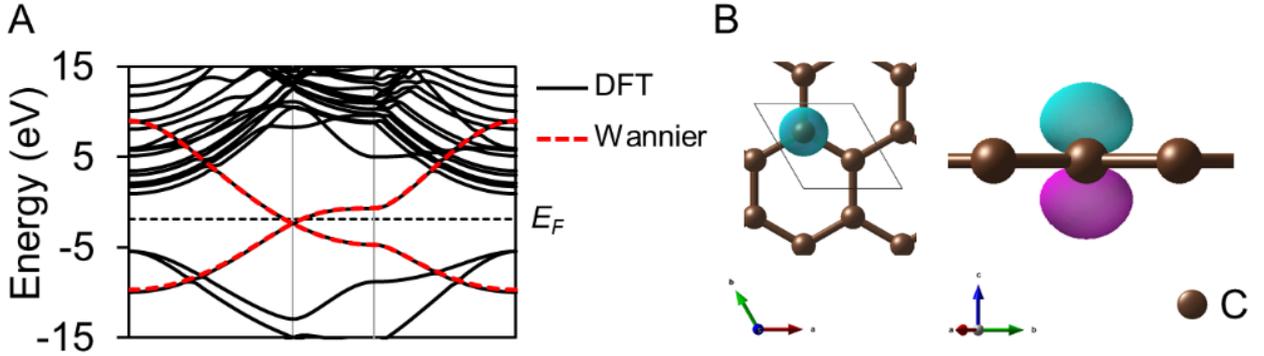

**Fig. 3. Graphene band structure and constructed Wannier orbitals.** (A) The graphene bands (black) from DFT and the obtained Wannier bands from the $p_z$ orbitals (red). (B) Wannier orbital for the lower energy band of the graphene $p_z$ orbitals in real space. The $p_z$ orbital is located around the carbon in the unit cell.

**Table 1.** Calculated onsite/hopping energy between corresponding sites, number of connecting sites, and distance between sites. $t_0$ denotes the on-site energy, and $t_1$, $t_2$, and $t_3$ denote the nearest, second nearest, and third nearest neighbor hopping of carbon $p_z$ orbitals, respectively.

| Onsite/hopping | Energy (eV) | Number of sites | Distance (Å) |
|---|---|---|---|
| $t_0$ | -1.994 | 1 | 0.00 |
| $t_1$ | -2.860 | 3 | 1.42 |
| $t_2$ | 0.236 | 6 | 2.47 |
| $t_3$ | -0.252 | 3 | 2.85 |

2. Calculation of band structure

We first execute quantum machine learning for a model with a $U$ of 0 eV to confirm that the Wannier band structure can be reproduced. In energy band calculations, the basis function adopted for the representation of wave functions is generally a one-electron basis function. For example, a linear combination of $|1000>, |0100>, |0010>,$ and $|0001>$ is used, and multiple-electrons basis functions such as $|0011>$ and $|0111>$ are not used. Thus, a penalty $\mu$ for the multiple-electrons basis is imposed on $H_{\mathbf{k}}$ in Eq. (6) [21]:

$$H'_k = H_k + \mu \left(\sum_{l\sigma} n_{kl\sigma} - 1\right)^2 \tag{11}$$

Quantum machine learning is performed for the eigenvalue problem of $H'_k$ at each k-point.

In Fig. 4A, the results of quantum machine learning for the π orbital of the graphene with a $U$ of 0 eV are shown as QML ($\mu$ = 15/50 eV) by dashed lines. For comparison, the band structure obtained by the exact diagonalization of $H_k$ with a $\mu$ of 15 eV is represented by gray lines (hereafter, "exact value"). At $\mu$ = 15 eV, QML values match the exact value at least near the Γ point, but deviate from the exact value significantly at around K point. At $\mu$ = 50 eV, QML values differ from the exact value at almost every k-point. The wavefunctions calculated with the exact diagonalization and QML using a $\mu$ value of 15 eV are both reasonably a linear combination of two basis functions (e.g., $\frac{1}{\sqrt{2}}(|0100> + |0001>)$) at the Γ point, while only one basis function (e.g., $|0100>$) appears in the calculated wavefunctions when the eigenvalues between the exact diagonalization and QML do not match (e.g., at around the Γ point using a $\mu$ of 50 eV, and the K point using $\mu$ = 15 eV). The results indicate that the desired eigenvalues cannot be obtained in QML if we adopt a large value for $\mu$. This problem is presumably caused by the difficulty in escaping from a broken-symmetry state such as $|0100>$ to a symmetric state such as $\frac{1}{\sqrt{2}}(|0100> + |0001>)$ because of a large $\mu$ value. The difficulty could be serious when the ability of expression for the wavefunctions by RBM is poor. In the remaining part, we adopted a $\mu$ value of 15 eV, which leads to a better result than $\mu$ = 50 eV.

To avoid the trapping problem of wavefunctions onto a broken symmetry state, we apply the transfer learning method to the electronic structure calculation of periodic systems. In a previous study on quantum machine learning with transfer learning [8], RBM parameters were optimized for the LiH molecule with a certain Li-H bond length (e.g., 1.00Å) by QML energy calculation, and these optimized RBM parameters were then used (i.e., transferred) as the initial parameters in the next QML calculation for the LiH molecule with a different Li-H bond length (e.g., 1.05Å). In this study, QML is first performed at the Γ point, and the obtained RBM parameters are used as the initial values of QML for the next k-point close to the Γ point. For the QML calculations at other k-points, the initial guess of the RBM parameters is transferred (i.e., copied) from the result of the QML calculation at the last step executed for the nearest k-point. By using the optimized RBM parameters at each k-point, wavefunctions will not only escape more easily from the broken symmetry states but they will also be optimized faster than optimization without transferring the RBM parameters. Hereafter, quantum machine learning using the transfer learning method will be referred to as QTL, and quantum machine learning without the transfer learning will be referred to as QML. QTL results are represented by the red dotted lines in Fig. 4A, and the difference between QTL and the exact diagonalization is shown in Fig. 4B. Since QTL calculations are executed from the Γ point on the left-hand side of Fig. 4B, we can confirm a rapid decrease in the difference when the parameter update and QTL calculation are repeated only for a few times. Note that the small difference at the K point originated from the degeneration of the ground state and the broken symmetry state, because these states take the same eigenvalues. The number of iterations in QTL for parameter optimization at each k-point also decreased by more than one order of magnitude (the number of QTL iterations is $\sim 10^2$, as compared to $\sim 10^3$ for QML). Thus, quantum machine learning can be employed for band structure calculations of periodic systems by taking the parameter optimization process into account (i.e., transfer learning). We also confirmed that when s(x) is limited to a real number, which is the same framework used in a previous study [8], the calculated band structure cannot be the same as the exact value, as indicated by black dotted lines in Fig. 4A; the value near the M point differs by about 1 eV from the result obtained using an imaginary number for s(x). Since the Hamiltonian is represented not for an isolated system but for a periodic system, the wavefunction has to be a complex number, and thus, the deviation obtained by using a real s(x) is reasonable.

In what follows, we execute the calculation result for a model with a $U$ of 9.3 eV to show the band splitting of upper and lower Hubbard bands. For a $U$ value of 0 eV, the up spin and down spin bands are energetically degenerated at every k-point; the up spin band is represented by the linear combination of $|1000>$ and $|0010>$, and the down spin band, by the linear combination of $|0100>$ and $|0001>$. Once we apply a nonzero $U$, the up and down spin bands split into two bands, the lower and upper Hubbard bands. Taking the average number of electrons in Eq. (6) as $<n\uparrow> = 1$ and $<n\downarrow> = 0$, we can obtain the up spin band as the lower Hubbard band. To calculate both lower and upper Hubbard bands, one more penalty term is added in the calculation for the upper Hubbard band so that the occupation state of the down spin site becomes the lowest eigenvalue. On the other hand, the same Hamiltonian as in Eq. (11) is used for the lower Hubbard band calculation.

$$H_{upper} = H_{\mathbf{k}} + \mu \left\{ \left( \sum_{l\sigma} n_{kl\sigma} - 1 \right)^2 + \left( \sum_{l} n_{kl\downarrow} - 1 \right)^2 \right\} \tag{12}$$

Figure 4C shows the band obtained by QML and QTL calculations for a $U$ value of 9.3 eV. The bands calculated with QML agree well with the exact values especially near the Γ point, as in the case of $U = 0$ eV. On the other hand, the bands calculated with QTL agree well with the exact values for every k-point. The Hubbard bands calculated with QML deviated significantly from the exact values because of the trapping problem, and the deviation of the bands was more noticeable in the upper Hubbard band than in the lower Hubbard band. This may be because the energy gap between the most stable and second most stable states in $H_{upper}$ (5 eV at Γ) is smaller than that in $H_{lower}$ (9 eV at Γ). Another possible reason is that $H_{upper}$ is imposed by a larger penalty term than $H_{lower}$, and thus, escaping from the broken symmetry states or other eigenvalues may be more difficult in the upper Hubbard band than in the lower one. It is expected that the problem will be solved by the same procedure as in the calculations with $U = 0$ eV, i.e., transfer learning. Figure 4D shows the difference between QTL and the exact value for $U = 9.3$ eV. Since the errors in QTL results relative to the exact values are at most a few meV, band structures can be correctly calculated when the QTL calculations are initiated from a reliable minimum solution. The above results suggest that quantum machine learning is useful for the calculation of physical properties of effective models for periodic systems.

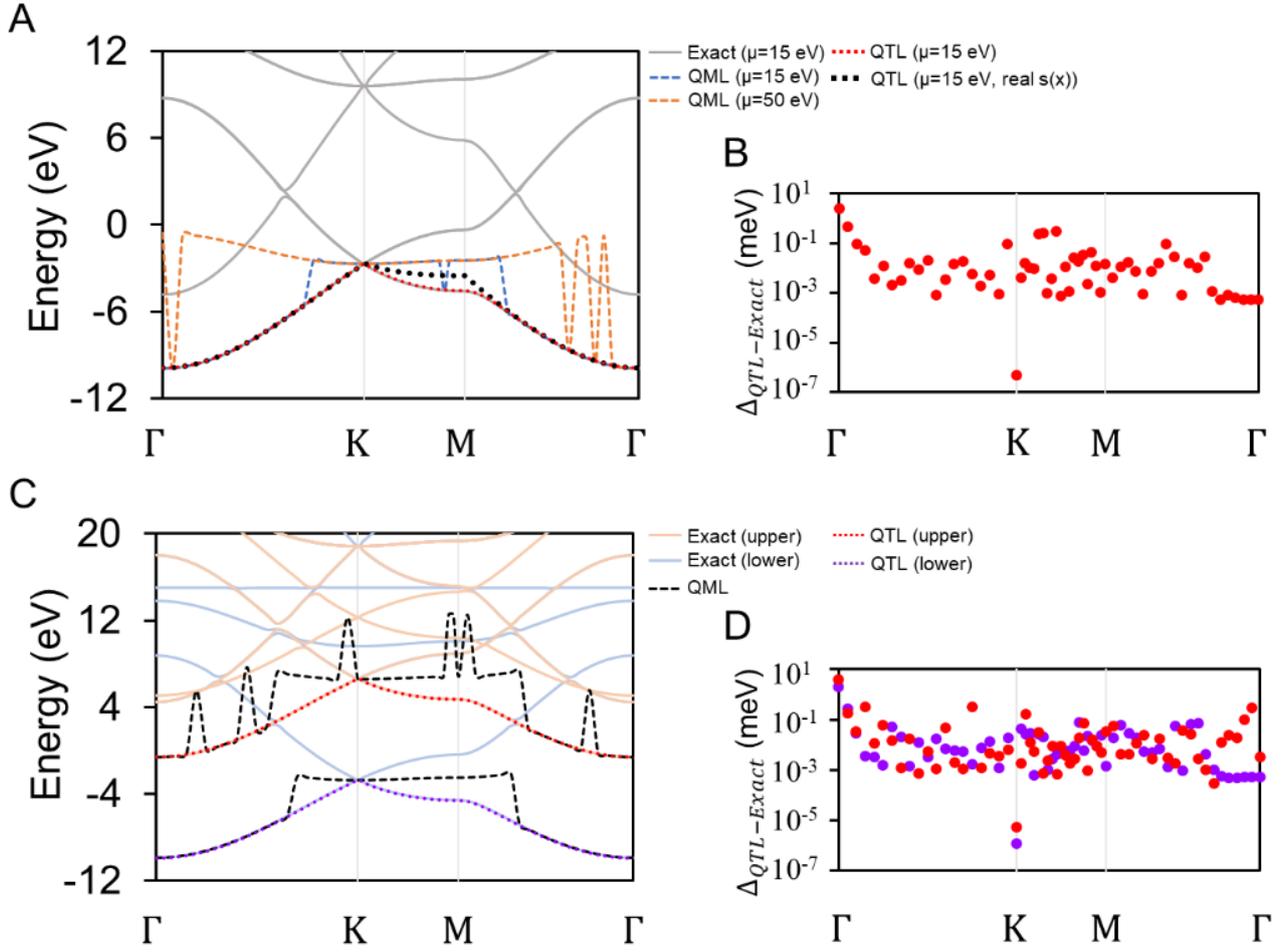

**Fig. 4. Band structures calculated by QML and QTL with $U$ values of 0 and 9.3 eV.** (A) The light gray, blue dashed, orange dashed, red dotted, and black dotted lines, represent, respectively, the results of exact diagonalization with $\mu = 15$ eV, QML with $\mu = 15$ eV, QML with $\mu = 50$ eV, QTL with $\mu = 15$ eV, and QTL by ignoring the imaginary part of $s(x)$. (B) Differences between exact values and QTL with $U = 0$ eV in logarithmic scale. QTL calculations are started at point Γ point, corresponding to the left side of the plot. (C) The light orange, light blue, black dashed, red dotted, and purple dotted lines are, respectively, the results of exact diagonalization with upper Hubbard band, exact diagonalization with lower Hubbard band, QML with both upper and lower Hubbard bands, QTL with upper Hubbard band, and QTL with lower Hubbard band. (D) Differences between exact values and QTL with $U = 9.3$ eV in logarithmic scale. Red and purple points are, respectively, the results with upper and lower Hubbard bands. QTL calculation procedure is the same as in Fig. 4B.

# III. CONCLUSION

In this study, we proposed a computational workflow to solve the many-body problems for periodic systems using quantum machine learning. The workflow comprised three steps: (i) global electronic structure calculation (DFT), (ii) effective model (Hubbard model) construction, and (iii) quantum machine learning. By reducing the number of orbitals and interactions, we can perform many-body calculations for materials even with small resources of quantum computers. In this study, we selected graphene as a test periodic system. After DFT and effective model construction, we calculated the Hubbard model of graphene with $U = 0$ eV using quantum machine learning. The obtained band dispersion agreed with the exact diagonalization (referred as "exact values"), especially at the k-points for which a large gap between eigenvalues was confirmed. Even for k-points with a small energy gap, an accuracy of a few meV was achieved by adopting the transfer learning procedure. In addition, quantum machine learning for the Hubbard model with a $U$ value of 9.3 eV under the mean-field approximation led to split bands by Hubbard $U$ (i.e., the upper and lower Hubbard bands), with accuracy relative to the exact values for $U = 9.3$ eV. Thus, quantum machine learning can be applied to electronic structure calculations of periodic models. Similarly to the history of classical computers, in the field of quantum chemical computation using the quantum algorithm, research mainly deals with organic molecules, rather than inorganic crystals. However, since machine power and algorithms have evolved dramatically even in the conventional computer, there is great potential for calculations involving inorganic crystals via hybrids between quantum and classical computers. We hope that the results obtained here will trigger quantum computational research on a wide range of materials.

## Acknowledgments

This work was supported in part by the KAKENHI (Grant-in-Aid for Scientific Research on Innovative Areas π-Figuration No. 26102017) to T.T. S.K. and T.T proposed the project. S.K. built the quantum machine learning simulator and carried out the study. S.K. and T.T. discussed the band structure calculations and wrote the paper.

# Appendix

We show the supplemental information about each part of the workflow.

1.Global electronic structure calculation

In this study, global electronic structure calculation is performed with density functional theory (DFT) implemented in the Quantum-Espresso package [22, 23]. The plane wave cutoff energy is set to 100 Ry, the k-point mesh is 16 × 16 × 1, and LDA the exchange-correlation functional and norm-conserving pseudopotential are used. The target material is graphene, which has two carbon atoms in the unit cell. The lattice constants are a=2.47Å in the in-plane direction, and 20 Å in the perpendicular direction. The crystal structure was visualized by using VESTA [24]. The in-plane lattice vectors $\boldsymbol{a_1}$ and $\boldsymbol{a_2}$, the carbon atom position vectors $\boldsymbol{r_A}$ and $\boldsymbol{r_B}$ in the unit cell, and the reciprocal lattice vectors $\boldsymbol{b_1}$ and $\boldsymbol{b_2}$ are as follows.

$$\boldsymbol{a_1} = a(1,0), \boldsymbol{a_2} = a\left(-\frac{1}{2}, \frac{\sqrt{3}}{2}\right)$$

$$\boldsymbol{r_A} = \frac{1}{3}\boldsymbol{a_1} + \frac{2}{3}\boldsymbol{a_2}, \boldsymbol{r_B} = \frac{2}{3}\boldsymbol{a_1} + \frac{1}{3}\boldsymbol{a_2}$$

$$\boldsymbol{b_1} = \frac{2\pi}{a}\left(1, \frac{1}{\sqrt{3}}\right), \boldsymbol{b_2} = \frac{2\pi}{a}\left(0, \frac{2}{\sqrt{3}}\right)$$

In the band calculation, the k-path of Γ (0,0) → K (1 / 3,1 / 3) → M (1 / 2,0) → Γ (0,0) is selected, and each section between the special k-points (e.g., from Γ to K) is divided by 20 grids.

2. Effective model construction

In this study, an effective model was created by the on-site term $t_0$ and hopping terms $t_m$ ($m$ = 1 - 3) listed in Table 1 and Hubbard $U$ in a previous study [14]. The maximally localized Wannier function and $t_m$ are obtained with the Wannier90 package [25] and shown in Table 1 of the main text.

$$H = t_0 \sum_{ij}^{A,B} \sum_{l} \sum_{\sigma}^{\uparrow,\downarrow} a_{ijl\sigma}^\dagger a_{ijl\sigma}$$

$$+t_1 \sum_{ij} \sum_{\sigma}^{\uparrow,\downarrow} (a_{ijB\sigma}^\dagger + a_{ij+1B\sigma}^\dagger + a_{i-1jB\sigma}^\dagger)a_{ijA\sigma} + (a_{ijA\sigma}^\dagger + a_{ij-1A\sigma}^\dagger + a_{i+1jA\sigma}^\dagger)a_{ijB\sigma}$$

$$+t_2 \sum_{ij}^{A,B} \sum_{l} \sum_{\sigma}^{\uparrow,\downarrow} (a_{i+1j+1l\sigma}^\dagger + a_{i+1jl\sigma}^\dagger + a_{ij-1l\sigma}^\dagger + a_{i-1j-1l\sigma}^\dagger + a_{i-1jl\sigma}^\dagger + a_{ij+1l\sigma}^\dagger)a_{ijl\sigma}$$

$$+t_3 \sum_{ij} \sum_{\sigma}^{\uparrow,\downarrow} (a^\dagger_{i-1j+1B\sigma} + a^\dagger_{i+1j+1B\sigma} + a^\dagger_{i-1j-1B\sigma}) a_{ijA\sigma}$$

$$+ (a^\dagger_{i+1j+1A\sigma} + a^\dagger_{i+1j-1A\sigma} + a^\dagger_{i-1j-1A\sigma}) a_{ijB\sigma}$$

$$+U \sum_{ijl} (n_{ijl\uparrow} <n_\downarrow> + <n_\uparrow> n_{ijl\downarrow} - <n_\uparrow><n_\downarrow>)$$

where $i, j$ are the lattice site indices, $l$ is atomic sites $A$ and $B$ in the lattice, $\sigma(=\uparrow,\downarrow)$ is the spin index, $a^\dagger/a$ is the creation/annihilation operator, and $n_{ijl\sigma}(= a^\dagger_{ijl\sigma} a_{ijl\sigma})$ is the number operator.

The Fourier transform is performed on the creation/annihilation operator.

$$H = t_0 \sum_{kl\sigma} a^\dagger_{kl\sigma} a_{kl\sigma}$$

$$+t_1 \sum_{k\sigma} \{e^{ik(r_B-r_A)}(1 + e^{ika_2} + e^{-ika_1}) a^\dagger_{kB\sigma} a_{kA\sigma} + e^{ik(r_A-r_B)}(1 + e^{ika_1} + e^{-ika_2}) a^\dagger_{kA\sigma} a_{kB\sigma}\}$$

$$+t_2 \sum_{kl\sigma} (e^{ika_1} + e^{-ika_1} + e^{ika_2} + e^{-ika_2} + e^{ik(a_1+a_2)} + e^{-ik(a_1+a_2)}) a^\dagger_{kl\sigma} a_{kl\sigma}$$

$$+t_3 \sum_{k\sigma} \{e^{ik(r_B-r_A)}(e^{ik(a_1+a_2)} + e^{-ik(a_1+a_2)} + e^{-ik(a_1-a_2)}) a^\dagger_{kB\sigma} a_{kA\sigma}$$

$$+ e^{ik(r_A-r_B)}(e^{ik(a_1+a_2)} + e^{-ik(a_1+a_2)} + e^{ik(a_1-a_2)}) a^\dagger_{kA\sigma} a_{kB\sigma}\}$$

$$+U \sum_{kl} (n_{kl\uparrow} <n_\downarrow> + <n_\uparrow> n_{kl\downarrow} - <n_\uparrow><n_\downarrow>)$$

$$= \sum_k \{\sum_{l,l'}^{A,B} \sum_\sigma^{\uparrow,\downarrow} \sum_m t_{kl'lm} a^\dagger_{kl'\sigma} a_{kl\sigma} + U \sum_l^{A,B} (n_{kl\uparrow} <n_\downarrow> + <n_\uparrow> n_{kl\downarrow} - <n_\uparrow><n_\downarrow>)\}$$

$$= \sum_k H_k$$

$r_{ijl} = ia_1 + ja_2 + r_l$, where $r_l = r_A$ or $r_B$. $t_{kl'lm}$ is the $m$-th neighboring hopping parameter between site $l'$ and site $l$ with the wave vector $k$. $H$ can be represented by the sum of the independent $H_k$.

3.Quantum machine learning

The update of the RBM parameters of $<A>$ ($=<H_k>$, $<H'_k>$ or $<H_{upper}>$) is executed in the following manner (see also supplementary information of [8]). Using the wavefunction $|\Psi> = \sum_x \sqrt{P(x)} s(x) |x>$, the expectation value $<A>$ is written as

$$<A> = \frac{<\Psi|A|\Psi>}{<\Psi|\Psi>} = \frac{\sum_{x',x} \overline{\left(\sqrt{P(x')}\, s(x')\right)} <x'|A|x> \sqrt{P(x)}\, s(x)}{\sum_x \left|\sqrt{P(x)}\, s(x)\right|^2}$$

where $P(x)$ and $s(x)$ are

$$P(x) = \frac{\sum_{\{h_j\}} e^{\sum_i a_i \sigma_i + \sum_j b_j h_j + \sum_{ij} w_{ij} \sigma_i h_j}}{\sum_{\{\sigma_i'\}} \sum_{\{h_j\}} e^{\sum_i a_i \sigma_i' + \sum_j b_j h_j + \sum_{ij} w_{ij} \sigma_i' h_j}}$$

$$s(x) = \tanh\left(\left(c + \sum_i d_i \sigma_i\right) + i\left(e + \sum_i f_i \sigma_i\right)\right).$$

The spin coordinate is represented by $x = \{\sigma_1 \ldots \sigma_4\}$, where the values of $\sigma_i$ or $\sigma_i'$ (visible) and $h_j$ (hidden) are 1 for spin up ↑ and -1 for spin down ↓. The summation with respect to $\sigma_i'$ and $h_j$ in $P(x)$ are taken for all the configuration of the visible and hidden layers. The gradients of $<A>$ are written as

$$\partial_{p_k} <A> = (<B(x;p_k)>_{x',x} + c.c) - <A>(<B(x;p_k)>_x + c.c),$$

where

$$p_k = a_i, b_j, w_{ij}, c, d_i, e, f_i$$

$$<B(x;p_k)>_{x',x} \equiv \frac{\sum_{x',x} \overline{\left(\sqrt{P(x')}\, s(x')\right)} <x'|A|x> B(x;p_k)\sqrt{P(x)}\, s(x)}{\sum_x \left|\sqrt{P(x)}\, s(x)\right|^2}$$

$$<B(x;p_k)>_x \equiv \frac{\sum_x B(x;p_k) \left|\sqrt{P(x)}\, s(x)\right|^2}{\sum_x \left|\sqrt{P(x)}\, s(x)\right|^2}$$

$$B(x;a_i) = \frac{1}{2}\sigma_i$$

$$B(x;b_j) = \frac{1}{2}\tanh(\theta_j)$$

$$B(x;w_{ij}) = \frac{1}{2}\tanh(\theta_j)\sigma_i$$

$$B(x;c) = \left(\frac{1}{s(x)} - s(x)\right)$$

$$B(x;d_i) = \left(\frac{1}{s(x)} - s(x)\right)\sigma_i$$

$$B(x;e) = i\left(\frac{1}{s(x)} - s(x)\right)$$

$$B(x;f_i) = i\left(\frac{1}{s(x)} - s(x)\right)\sigma_i$$

$$\theta_j = \sum_i w_{ij}\sigma_i + b_j.$$

$\sigma_i$ is related only on $i$-th spin of $x$, and therefore the value of $\sum_i w_{ij}\sigma_i$ in $\theta_j$ is determined by the whole configuration of $x$. For example, if $x$ is $\{\uparrow\downarrow\downarrow\uparrow\}$, $(\sigma_0,\sigma_1,\sigma_2,\sigma_3) \to (1,-1,-1,1)$ and $\sum_i w_{ij}\sigma_i \to w_{0j} - w_{1j} - w_{2j} + w_{3j}$.

Here we define a vector $\boldsymbol{v}_l$ as the RBM parameters updated in the $l$-th iteration, and $\boldsymbol{v}_l$ can be written as

$$\boldsymbol{v}_l = (a_0, \dots, a_{m-1}, b_0, \dots, b_{n-1}, w_{00}, \dots, w_{m-1\,n-1}, c, d_0, \dots, d_{m-1}, e, f_0, \dots, f_{m-1}),$$

where $m$ and $n$ are respectively the numbers of the unit in the visible and hidden layers, and we omit the index $l$ for each RBM parameters in the above notation. For example, $a_0$ indicates the $l$-th updated $a_0$, and so on.

We also define $\boldsymbol{g}_l$ to represent the gradient of $<A>$ with respect to the RBM parameters in $l$-th iteration as,

$$\boldsymbol{g}_l = (\partial_{a_0}<A>, \dots, \partial_{w_{m-1\,n-1}}<A>, \partial_c <A>, \dots, \partial_{f_{m-1}}<A>).$$

The $\boldsymbol{v}_l$ is updated based on Adam [18] by using $\boldsymbol{g}_l$, that is,

$$\boldsymbol{v}_{l+1} = \boldsymbol{v}_l - \left(\frac{\alpha}{\widehat{\boldsymbol{V}_{l+1}}^{\frac{1}{2}} + \epsilon}\right)\widehat{\boldsymbol{M}_{l+1}}$$

$$\widehat{\boldsymbol{M}_{l+1}} = \frac{\boldsymbol{M}_{l+1}}{1 - \beta_1^{l+1}}$$

$$\widehat{\boldsymbol{V}_{l+1}} = \frac{\boldsymbol{V}_{l+1}}{1 - \beta_2^{l+1}}$$

$$\boldsymbol{M}_0, \boldsymbol{V}_0 = 0$$

$$\boldsymbol{M}_{l+1} = \beta_1 \boldsymbol{M}_l + (1-\beta_1)\boldsymbol{g}_l$$

$$\boldsymbol{V}_{l+1} = \beta_2 \boldsymbol{V}_l + (1-\beta_2)\boldsymbol{g}_l^2.$$

Learning rate $\alpha$ is 0.01 and the other hyperparameters are $\beta_1 = 0.9$, $\beta_2 = 0.999$, $\epsilon = 10^{-8}$. The initial value of each RBM parameters are determined by random numbers ranging from -0.02 to 0.02, and the iteration number $l$ is started from zero. Convergence threshold of $<H_k>$ and $<H_{upper}>$ variation was $10^{-6}$ eV.